\documentclass[fleqn,12pt,twoside]{article}
\usepackage{espcrc1}

\usepackage{graphicx}
\usepackage[figuresright]{rotating}

\newcommand{\AmS}{{\protect\the\textfont2
  A\kern-.1667em\lower.5ex\hbox{M}\kern-.125emS}}

\title{Precursor of Color Superconductivity}

\author{M. Kitazawa\address[KYOTO]{Department of Physics,
 Kyoto University, Kyoto 606-8502, Japan},
T. Koide\address[YITP]{Yukawa Institute for Theoretical Physics,
Kyoto University, Kyoto 606-8502, Japan},
T. Kunihiro\addressmark[YITP] 
and 
Y. Nemoto\address[BNL]{RIKEN BNL Research Center, BNL, Upton, NY 11973}}

\begin{document}

\maketitle

\begin{abstract}
We investigate possible precursory phenomena of color superconductivity
at finite temperature $T$ with an effective theory of QCD.
It is found that the fluctuation of the diquark pair field 
exists with a prominent strength
even well above the critical temperature $T_c$. 
We show that such a fluctuaiton forms a collective mode, 
the corresponding pole of which 
approaches the origin as $T$ is lowered to $T_c$ 
in the complex energy plane.
We discuss the possible relevance of the precursor to
the observables to be detected in heavy-ion collisions.
\end{abstract}

\section{INTRODUCTION}

Determining the phase structure of Quantum Chromodynamics (QCD) is 
one of the central problems in hadron physics.
In particular, at low temperature ($T$) 
and non-zero chemical potential ($\mu$), 
one may expect that quark matter has a Fermi surface, 
which gives rise to an interesting complication;
if there is some attractive channel in the quark-quark interaction, 
the Fermi surface becomes unstable and quarks form 
Cooper pairs, leading to 
a color superconductivity (CS) with  the color-gauge symmetry broken
\cite{ref:REV}.

Since the critical temperature ($T_c$) of 
CS is expected to be small,
it would be difficult to create the color-superconducting quark matter 
by heavy-ion collisions.
So it is natural that  people have exclusively 
considered  neutron stars as a physical system to which 
CS may have a relevance.

However, there is a possibility to obtain information on CS 
by heavy-ion collisions through possible {\em fluctuations} 
of the pair field.
In fact, it is known that  
in the ordinary superconductivity in metals,
there exist fluctuations of  Cooper pairs 
even above the critical temperature where the pair
condensate does not exist:
Such fluctuations affect various transport coefficients, 
which are experimentally observed\cite{ref:Tsuneto}.
In the usual superconductivity in metals, 
the pairing interaction due to the phonon exchange is so weak that 
the mean field approximation works well except in the immediate
vicinity of the critical point.
In other words, 
the fluctuations become important only near the critical temperature.
However, in the case of CS at intermediate densities, 
large fluctuations may survive even well above $T_c$
owing to the strong coupling which may be described by
a low-energy effective  Lagrangian of QCD.
If there are large fluctuations even well above $T_c$, 
heavy-ion collisions on Earth may have a possibility to be utilized 
to study the physics of CS through 
examining precursory phenomena of CS.

\begin{figure}
\begin{minipage}{8cm}
\includegraphics{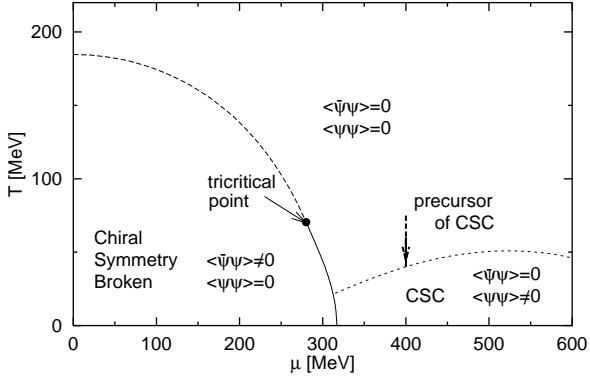}
\caption{The calculated phase diagram in $T$-$\mu$ plane 
in our model.
The Solid and dashed line denote the critical line 
 of  a first and second order phase transition, 
respectively.}
\end{minipage}
\end{figure}

\begin{figure}[h]
\vspace{-9cm}\hspace*{9cm}
\begin{minipage}{6cm} 
\begin{center}
\includegraphics[width=4cm]{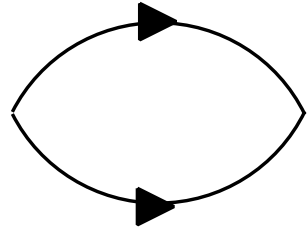}
\caption{The one loop diagram of the particle-particle correlation.
}
\end{center}
\end{minipage}
\end{figure}

\vspace{1.5cm}

\section{RESPONSE OF PAIR FIELD}

We use the two-flavor Nambu-Jona-Lasinio model in the chiral limit,
\cite{ref:KKKN}
\begin{eqnarray}
{\cal L} &=& \bar{\psi}i/\hspace{-2mm}\partial \psi 
+ G_{S}[(\bar{\psi}\psi)^2
+ (\bar{\psi}i\gamma_{5}\vec{\tau} \psi)^2] 
+ G_{C}(\bar{\psi}i\gamma_{5}\tau_{2}
\lambda_{2}\psi^{C})(\bar{\psi}^{C}i\gamma_{5}\tau_{2}\lambda_{2}\psi),
\end{eqnarray}
where $\psi^{C} \equiv C\bar{\psi}^{T}$ 
with $C = i\gamma^{2}\gamma^{0}$ being the charge conjugation
operator, and 
$\tau_{2}$ and $\lambda_{2}$ are the second component of the 
Pauli and Gell-Mann matrices representing 
the flavor SU(2) and color SU(3), respectively.

The phase structure of our model in the mean field approximation 
is shown in Fig.~1.
The dashed line denotes the critical line
of the second order phase transition 
and the solid line first order phase transition.
The tricritical point where the order of the chiral transition changes
from the second to first order is located at about $(T,\mu)=(70,280)$~MeV.

Now that the phase structure is determined,
we investigate the $T$ dependence of the 
fluctuation of the diquark pair field in the Wigner phase 
above $T_c$;
we calculate the response of 
the pair field to an external one in the linear response theory.
When an external field $\Delta^{*}_{ex}$ is added in the Wigner phase, 
the induced pair field $\Delta^{*}_{ind}$ is given by
\begin{eqnarray}
\Delta^{*}_{ind}({\bf k},\omega_{n})
=
\frac{-G_{C}{\cal Q}({\bf k},\omega_{n})}{1+G_{C}{\cal Q}({\bf k},\omega_{n})}
\Delta^{*}_{ex} 
\equiv {\cal D}({\bf k},i\omega_{n})\Delta^{*}_{ex}, \label{eqn:res}
\end{eqnarray}
where ${\cal Q}({\bf k},\omega_{n})$ 
is a particle-particle correlation function 
corresponding diagrammatically to Fig.~2.
The function ${\cal D}({\bf k},i\omega_{n})$ 
is still a function of imaginary time.
By carrying out the analytic continuation,
we then obtain, what we call, the response function 
$D^{R}({\bf k},\omega)$.
Here it should be noticed that the following equality holds 
\begin{eqnarray}
\label{eqn:critical}
\left.1+G_{C}{\cal Q}({\bf 0},0)\right|_{T=T_{c}}=0,
\end{eqnarray}
on account of
the self-consistency condition for the second-order phase transition 
of the diquark condensate 
at $T=T_c$. This equality is the origin to lead to 
the various precritical phenomena for CSC.
The nature of the fluctuation to be discussed below 
will not be altered even when the
phase transition is {\it weak} first order owing to the
fluctuations of the gluon fields.

\begin{figure}[t]
\begin{minipage}{8cm} 
\includegraphics[width=9cm]{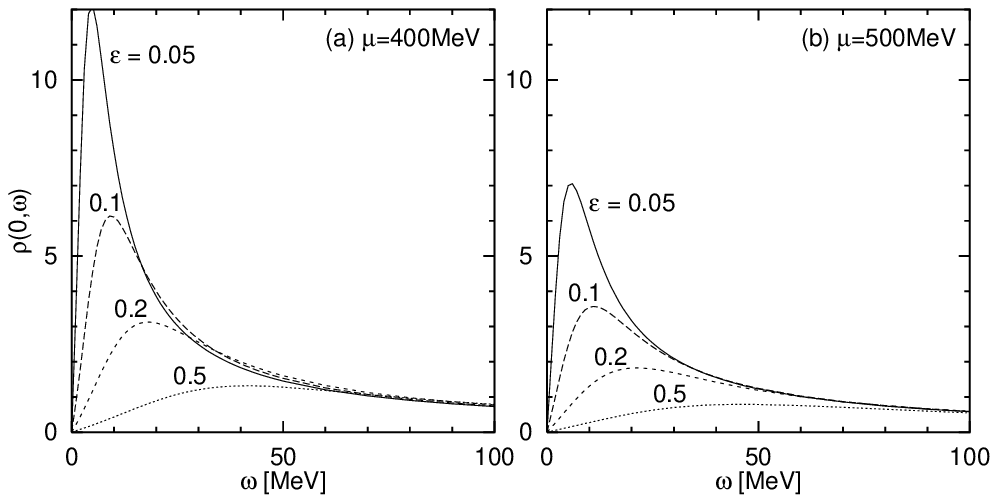} 
\caption{The spectral function  for the pair field at $T>T_c$ 
with $\epsilon \equiv (T-T_c)/T_c = 0.05$,$0.1$,$0.2$ and $0.5$ 
at $\mu = 400$MeV (a) and $\mu = 500$MeV (b).
}
\end{minipage}
\end{figure}

\begin{figure}[t]
\vspace{-9cm}\hspace*{9cm}
\begin{minipage}{7cm}
\begin{center} 
\includegraphics[width=5.5cm]{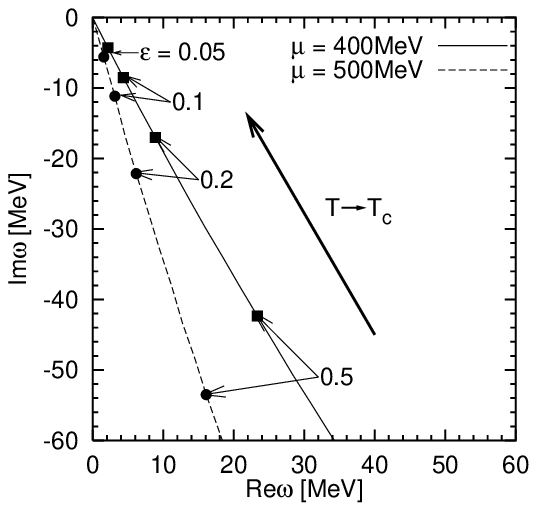} 
\caption{The spectral function  for the pair field at $T>T_c$ 
with $\epsilon \equiv (T-T_c)/T_c = 0.05$,$0.1$,$0.2$ and $0.5$ 
at $\mu = 400$MeV (a) and $\mu = 500$MeV (b).
}
\end{center}
\end{minipage}
\end{figure}

The strength of the fluctuation
is given by the spectral function
\begin{eqnarray}
\rho({\bf k},\omega)= \frac{1}{2\pi G_{C}}{\rm Im}D^{R}({\bf k},\omega),
\end{eqnarray}
where $\omega >0$.
The $T$ dependence of the spectral function 
at $\mu=400$ MeV and zero momentum transfer (${\bf k}=0$) 
is shown in Fig.~3(a). One can see a prominent peak 
moving toward the origin grows as $T$
is lowered toward $T_c$ as indicated by the arrow in Fig.~1.
In the experimental point of view, 
it is interesting that the peak survives at $T$ even well 
above $T_c$ with $\epsilon\equiv (T-T_c)/T_c\sim 0.2$.
This means that the precritical region of CSC is one to two 
order larger in the unit of $T_c$ than that in 
the usual electric superconductors of metals.
The result  for  $\mu=500$ MeV  is 
shown in Fig.~3(b);  one sees that although 
the growth of the peak becomes relatively moderate, 
the qualitative features do not change.

\section{SOFT MODE AND PARACONDUCTIVITY}

The existence of the peak with the narrow width suggests 
an existence of a well-defined collective mode as an
elementary excitation in the Wigner phase.
This can be examined by searching possible poles of 
the response function.
The existence of a pole 
$\omega=\omega(k)$ with given ${k}$ means that 
the pair field $\Delta_{ind}({k},\omega(k))$ can be created 
even with an infinitesimal external field $\Delta^{*}_{ex}$ 
as seen from Eq.~(\ref{eqn:res});i.e., 
the system admits  spontaneous
excitation of a collective mode 
with the dispersion relation $\omega=\omega(k)$.
(One should notice that 
the gap equation  Eq.~(\ref{eqn:critical})  
implies that $z=0$ is a solution at $T=T_c$, i.e., 
there exists a zero mode at the critical point.)
The singular behavior seen in the spectral function
near the critical point in Fig.~3 is caused by the presence of the zero mode.
A numerical calculation shows that 
there indeed exists a pole in the lower half plane at  $T>T_c$.
Fig.~4 shows how the pole moves as $T$ is lowered toward $T_c$ for
$\mu=400$ and $500$ MeV at ${\bf k}=0$.
One sees that the pole approaches the origin 
as $T$ is lowered toward $T_c$.
Such a mode whose energy tends to vanish as the system approaches the 
critical point of a phase transition is called a soft mode.
Another characteristic feature of the soft mode is that 
the absolute value of the imaginary part $\omega_i$ of the pole 
is larger than that of the real part $\omega_r$ for both $\mu$.
This feature may have an important implication for an effective
equation describing the dynamical phase transition for CSC:
The dynamical behavior of the order parameter near $T_c$
is well described by a non-linear diffusion equation like 
the time-dependent Ginzburg-Landau equation.

\begin{figure}[t]
\begin{minipage}{15cm} 
\begin{center}
\includegraphics[width=15cm]{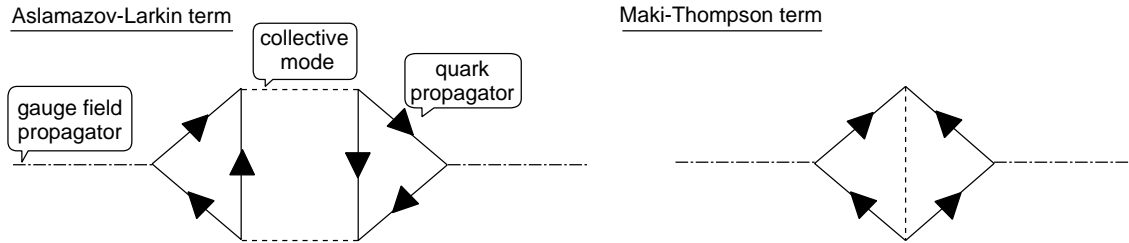} 
\caption{Diagrams which contribute paraconductivity.
The left and right figures correspond to the Aslamazov-Larkin and 
Maki-Tompson terms, respectively.
}
\end{center}
\end{minipage}
\end{figure}

Such a soft mode gives rise to anomalous behaviors of 
several transport coefficients; for example, 
an anomalous excess of the 
electric conductivity above $T_c$ which is known as 
paraconductivity (PC)\cite{ref:AL}.
The diagrammatic contributions to PC are 
known as  Aslamazov-Larkin and Maki-Tompson terms, which are shown in 
Fig.~5.
It is worth emphasizing that these diagrams can be 
regarded as a modification of 
the self-energy of  gauge fields, i.e., the gluons
and the photons, which
 decay into dilepton pairs.
This implies that photons and hence dileptons from the system
can carry some information of the fluctuation of CS.
We also notice that the pairing fluctuations can give
rise to a pseudo-gap phenomena known in high-$T_c$ superconductivity
of copper-oxides. It would be intriguing to pursue possible
similarity of CSC and  high-$T_c$ superconductivity.
These are under investigations.


\begin{thebibliography}{99}
%
\bibitem{ref:REV}
For recent reviews, see K.~Rajagopal and F.~Wilczek, Chapter 35 in the Festschrift in honor of B. L. Ioffe, 
"At the Frontier of Particle Physics / Handbook of QCD", 
M. Shifman, ed., (World Scientific);
M.~Alford, Ann. Rev. Nucl. Part. Sci. {\bf 51}, 131 (2001). 
%
\bibitem{ref:Tsuneto}
T.~Tsuneto, ``Superconductivity and superfluidity'', 
Cambridge University Press, 1998.
%
\bibitem{ref:KKKN}
M.~Kitazawa, T.~Koide, T.~Kunihiro and Y.~Nemoto, 
Phys.~Rev.~{\bf D65}, 091504(R) (2002).
%
\bibitem{ref:AL}
L.~G.~Aslamazov and A.~I.~Larkin,
Sov.~Phys.~Solid~State {\bf 10},
875 (1968);\, 
K.~Maki, Prog.~Theor.~Phys. {\bf 40}, 193 (1968);\,
R.~S.~Thompson, Phys.~Rev. {\bf B1}, 327 (1970).
\end{thebibliography}
\end{document}